\documentclass[useAMS,usenatbib]{mn2e}

\usepackage{epsf}
\usepackage{amsmath}

\newcommand{\hmpc}{h^{-1}{\rm Mpc}}

\newcommand{\asf}{\alpha_{\rm sf}}

\newcommand{\gad}{{\sc Gadget-2}}
\newcommand{\ion}[2]{\hbox{#1\,{\sc #2}}}

\title[The Nature of Sub-mm Galaxies]{The Nature of Sub-millimetre Galaxies in Cosmological Hydrodynamic Simulations}

\author[R. Dav\'e et al.]{
\parbox[t]{\textwidth}{\vspace{-1cm}
Romeel Dav\'e$^1$, Kristian Finlator$^{1,2}$, Benjamin D. Oppenheimer$^{1,3}$, 
Mark Fardal$^4$, Neal Katz$^4$, Du\v{s}an Kere\v{s}$^5$, David H. Weinberg$^6$}
\\\\$^1$ Astronomy Department, University of Arizona, Tucson, AZ 85721
\\$^2$ Physics Department, University of California, Santa Barbara, CA 93106
\\$^3$ Leiden Observatory, P.O. Box 9513, NL-2300 RA Leiden, The Netherlands 
\\$^4$ Astronomy Department, University of Massachusetts, Amherst, MA 01003
\\$^5$ Harvard-Smithsonian Center for Astrophysics, Cambridge, MA 02138
\\$^6$ Astronomy Department, Ohio State University, Columbus, OH 43210
}

\begin{document}

\pubyear{2009}

\maketitle

\label{firstpage}

 \begin{abstract}
We study the nature of rapidly star-forming galaxies at $z=2$ in
cosmological hydrodynamic simulations, and compare their properties to
observations of sub-millimetre galaxies (SMGs).  We identify simulated
SMGs as the most rapidly star-forming systems that match the observed
number density of SMGs.  In our models, SMGs are massive galaxies sitting
at the centres of large potential wells, being fed by smooth infall and
gas-rich satellites at rates comparable to their star formation rates
(SFR).  They are not typically undergoing major mergers that significantly
boost their quiescent SFR, but they still often show complex gas morphologies
and kinematics.  Our simulated SMGs have stellar masses of $M_*\sim
10^{11-11.7} M_\odot$, SFRs of $\sim 180-500\; M_\odot/$yr, a clustering
length of $\sim 10\, h^{-1}$Mpc, and solar metallicities.  The SFRs are lower
than those inferred from far-IR data by $\sim\times3$, which we suggest
may owe to one or more systematic effects in the SFR calibrations.
SMGs at $z=2$ live in $\sim 10^{13} M_\odot$ halos, and by $z=0$ they
mostly end up as brightest group galaxies in $\sim 10^{14} M_\odot$ halos.
We predict that higher-$M_*$ SMGs should have on average lower specific
SFRs, less disturbed morphologies, and higher clustering.  We also predict
that deeper far-IR surveys will smoothly join SMGs onto the massive end
of the SFR$-M_*$ relationship defined by lower-mass $z\sim 2$ galaxies.
Overall, our simulated rapid star-formers provide as good a
match to available SMG data as merger-based scenarios, 
offering an alternative scenario that emerges naturally
from cosmological simulations.
\end{abstract}

 
\section{Introduction}\label{sec: intro}

Sub-millimetre galaxies \citep[SMGs;][]{bla02} are among the most
enigmatic objects in the Universe.  SMGs were originally identified using
the Submillimetre Common-User Bolometer Array (SCUBA) at 850$\mu m$ and
450$\mu m$ on the James Clerk Maxwell Telescope, with multi-wavelength
follow-up from the radio to the X-rays.  Their most remarkable
property is their enormous bolometric luminosity, sometimes exceeding
$10^{13}L_\odot$, most of which is emitted as dust-reprocessed light in
the far infrared (IR).  Because the sub-millimetre band probes a falling
(versus wavelength) portion of the galaxy's spectral energy distribution
(SED), SMGs have the fortuitous property that their apparent brightness is
relatively invariant over a large redshift range.  In principle, this can
be exploited to probe star formation and/or black hole growth over a large
fraction of cosmic time, but in practice this requires ancillary data
to both locate the object precisely and to obtain its redshift.  Radio
interferometry has proven useful for this, enabling precise location of
an optical counterpart for which a redshift can be obtained.  In this way,
the majority of SMGs are found to lie at $z\sim 1.5-3$~\citep{are07}, with
the most distant one currently known at $z=4.76$~\citep{cop09}.  However,
the sensitivity of current radio interferometers dies out at $z\ga 3$,
meaning that such a selection may bias the overall redshift distribution
to lower redshifts.  Recent data using AzTEC at 1.1~$\mu$m suggests a
somewhat higher mean redshift for these sources than the 850-$\mu$m
selected ones by including 24$\mu$m counterparts~\citep{chapin09}.
In any case, it is clear that SMGs probe the brightest galaxies during
the most active cosmic epoch for galaxy assembly.

Observational studies of SMGs are among the fastest-growing areas of
extragalactic astronomy.  Uniform and complete samples of SMGs are
now being obtained such as the SCUBA Half-Degree Extragalactic Survey
\citep[SHADES;][]{cop06} covering 0.5~square degrees to a depth of
2~mJy, and there are other large samples such as that of \citet{cha05}.
Such surveys can in principle provide interesting discriminants between
models for the origin of SMGs~\citep{vank05}.  With new instruments such
as LABOCA on the Atacama Pathfinder Experiment, AzTEC, and eventually
SCUBA-2 on JCMT, and upcoming facilities in the near future such as
the Large Millimeter Telescope, catalogues of SMGs will continue to
improve rapidly.

Despite rapidly accumulating data, the nature of the SMGs remains
poorly understood.  Locally, the most bolometrically luminous galaxies
are ultraluminous IR galaxies with luminosities of $L_{IR}>\sim
10^{12}L_\odot$~\citep[ULIRGs;][]{san96}, and are seen to be ongoing major
gas-rich mergers, e.g. Arp~220.  By the First Law of Ducks\footnote{\tt http://www.bpd411.org/duck.html}, it was (and
often continues to be) assumed that SMGs are analogous merger events at
high redshift, simply scaled up to larger luminosities owing to the higher
gas content of early galaxies.  The major merger hypothesis is supported
by the preponderance of close neighbours~\citep[e.g.][]{ivi07} and their
often disturbed morphologies~\citep{men07,tac08}.  Although recent
models suggest a strong coevolution of stellar and black hole growth
in major mergers~\citep[e.g.][]{dim05}, observations indicate that star
formation dominates the bolometric luminosity, and active galactic nuclei
(AGN) provide only a minor contribution~\citep{ale05,men07,pop08,cle08}.
This isn't necessarily a difficulty for the merger scenario if the peak
star formation occurs at an earlier phase in the merger relative to peak
black hole growth, as such models suggest~\citep[e.g.][]{nar09a}.  But it
does mean that the far-IR luminosity of SMGs can be plausibly translated
into a star formation rate (SFR) without the fear of large contamination
by AGN.  If one adopts local calibrations~\citep[SFR$=4.5\times 10^{-44}
L_{\rm FIR}$ erg/s;][]{ken98a}, the inferred SFRs are typically many
hundreds to several thousands of solar masses per year.

Hierarchical models of galaxy formation have traditionally had difficulty
reproducing the observed numbers and fluxes of SMGs.  Two theories 
have figured prominently in recent discussions, both exploiting the large
enhancement in star formation rates believed to accompany gas-rich major
mergers: (1) SMGs are large, gas-rich, merger-induced starbursts, caught
in a special phase where their luminosity is significantly enhanced
over their quiescent state~\citep[e.g.][]{nar09a} and (2) SMGs are
modest-sized merger-induced starbursts whose bolometric luminosity
is greatly enhanced by an extremely top-heavy stellar initial mass
function~\citep[IMF;][]{bau05}.

The first scenario, large major mergers, is the canonical one, and
it has had recent success reproducing the detailed SEDs and CO
properties of SMGs from merger simulations~\citep{cha08,nar09a,nar09b}.
However, owing to the rarity of such large galaxies at high-$z$ and
the short duration of their merger-induced boost, this scenario may
have difficulty reproducing the number density of SMGs within a
hierarchical context.  To be quantitative, to produce SMG
fluxes ($\sim 5$~mJy), \citet{nar09a} needed to merge galaxies with
stellar masses of $\sim 3\times 10^{11}M_\odot$.  Such galaxies are
above $L_*$ at these epochs and at $z\sim 2$ the stellar mass
function of \citet{mar09} and \citet{kaj09} indicates that their
number density is $\approx 5\times 10^{-5}$~Mpc$^{-3}$ (comoving).
In addition, the dimensionless major merger rate for such objects is
around unity per Hubble time at $z\sim 2-3$~\citep[even allowing
for up to $3:1$ mergers;][]{guo08,hop09}, meaning that if such
objects are observable as SMGs for 100~Myr~\citep[as indicated in
Figure~1 of][]{nar09a}, then at $z=2$ the predicted number density of 
SMGs would be $\sim 2\times 10^{-6}$~Mpc$^{-3}$.  The observed
number density is up to an order of magnitude higher than this, 
$1-2\times 10^{-5}$~Mpc$^{-3}$ \citep{bor05,swi06,dye08}.
Similar merger simulations by
\citet{cha08} found that even higher masses are required to produce
SMG fluxes, which would result in even lower abundances.  The stellar
mass functions at $z\sim 2$ are still subject to significant
systematics~\citep[e.g.][]{muz09,kaj09}, so the discrepancy could
be less.  \citet{gen08} found that there are enough halos with
sufficient accretion rates undergoing major mergers to account for
perhaps a quarter to a half of the observed SMGs, albeit with
some assumptions about how rapid galaxies merge compared to their
halo merger times.
\cite{dek09} have also proposed that rapidly star-forming
galaxies at high redshift are powered primarily by smooth accretion
and minor mergers, though they suggested that some SMGs themselves
might still be powered by major mergers.
Hence while large gas-rich mergers
can certainly produce objects that look like SMGs, it is possible that
a large fraction of SMGs are in fact not such objects.

An alternative scenario for SMGs was presented by \citet{bau05}, based
on semi-analytic galaxy formation models.  Owing to the paucity of
large major mergers in hierarchical models, \citet{bau05} were driven to
argue that more common mergers of sub-$L_*$ galaxies give rise to SMGs.
But such modest-sized mergers fell far short of reproducing the observed
SMG fluxes.  To boost the flux, they then invoked an IMF that is
flat above $1 M_\odot$ during the merger event.  This model can match both
the number density and far-IR fluxes of SMGs (although the actual SFRs
are well below those inferred using standard conversion factors), at the
cost of introducing a radically top-heavy IMF in such systems.  However,
subsequent data on SMGs has not supported this model.  For instance,
this scenario predicts stellar masses significantly lower than that
recently inferred for SMGs from near-IR data~\citep[e.g.][]{swi08}.
Also, the IMF directly inferred from CO and dynamical observations of SMGs
disfavours such dramatic departures from a standard \citep[e.g.][]{cha03}
IMF~\citep{tac08}.  \citet{swi08} discusses some possible ways to
reconcile these discrepancies, but a fully consistent scenario has yet to
be put forth.

A third, less discussed scenario for SMGs was presented in \citet{far01}.
This work considered the nature of SMGs in cosmological hydrodynamic
simulations, and showed that for reasonable values of dust temperatures
such simulations can reproduce the observed 850$\mu$m number counts.
In these models, SMGs were galaxies forming stars at high rates,
$>100\; M_\odot/$yr, but were massive systems doing so on relatively long
timescales rather than in short merger-induced bursts.  A similar conclusion
was reached by \citet{fin06}.  In this work on the properties of $z=4$
galaxies from cosmological hydrodynamic simulations, it was noted that
a few galaxies had extremely high star formation rates, including two
exceeding $1000\; M_\odot/$yr.  These galaxies were not undergoing a large
burst of star formation owing to a major merger, but instead were massive
galaxies that had been forming stars at hundreds of $M_\odot/$yr for
some time.  Such high SFRs are expected at high redshifts because the
dense intergalactic medium (IGM) and short cooling times yield large
accretion rates~\citep{ker05,dek06,ker08}.  This scenario, therefore, suggests
that SMGs are not associated with major mergers at all, but are instead
super-sized versions of normal star-forming galaxies.

In this paper, we follow on the works of \citet{far01} and \citet{fin06}
to examine in detail the nature of simulated galaxies with high SFRs at
redshift $z\sim 2$, and how their properties compare to the wealth of
recently obtained multiwavelength data on SMGs.  We employ cosmological
hydrodynamic simulations including a heuristic implementation of galactic
outflows that has been shown to match observations of more typical
$z\sim 2$ galaxies reasonably well, described in \S\ref{sec: sim}.
In \S\ref{sec:highsfr} we study the stellar masses, star formation rates,
gas fractions, metallicities, environments, and clustering of simulated
SMGs, and show that the ``super-sized star-former" scenario comes
interestingly close to reproducing many of their observed properties.
In \S\ref{sec:examples} we study four simulated SMGs in more detail,
including their morphology, kinematics, and star formation and enrichment
histories.  A significant challenge to our scenario is that the SFRs
inferred for SMGs are factors of few higher than in our simulated SMGs
at a fixed number density.  In \S\ref{sec:imf} we suggest that SMG star
formation rates have been modestly overestimated, and place this idea
in a broader context of galaxy evolution at those epochs.  We summarise
and discuss the implications of our results in \S\ref{sec:summary}.

\section{Simulations} \label{sec: sim}

We employ our modified version of the N-body+hydrodynamic code \gad,
described more fully in \citet{opp08}; here we review the basic
ingredients.  \gad\ uses a tree-particle-mesh algorithm to compute
gravitational forces on a set of particles, and an entropy-conserving
formulation of Smoothed Particle Hydrodynamics \citep[SPH;][]{spr05}
to simulate pressure forces and shocks on the gas particles.  We
include radiative cooling from primordial~\citep[following][]{kat96}
and metal~\citep[based on][]{sut93} elements, assuming ionisation
equilibrium.  Star formation follows a \citet{sch59} Law calibrated
to the \citet{ken98} relation; particles above a density threshold
where sub-particle Jeans fragmentation can occur are randomly
selected to spawn a star with half their original gas mass.  The
interstellar medium (ISM) is modelled through an analytic subgrid
recipe following \citet{mck77}, including energy returned from
supernovae~\citep{spr03a}.

Chemical enrichment is followed in four individual species (C, O, Si,
Fe) from three sources: Type II SNe, Type Ia SNe, and stellar mass
loss from AGB stars.  The former instantaneously enriches star-forming
particles, whose metallicity can then be carried into the IGM via outflows
(described below).  Type~Ia modelling is based on the fit to data by
\citet{sca05}, as a prompt component tracing the star formation rate
and a delayed component (with a 0.7~Gyr delay) tracking stellar mass.
Stellar mass loss is derived from \citet{bru03} population synthesis
modelling assuming a \citet{cha03} IMF.  Delayed feedback adds energy and
metals to the three nearest gas particles; Type~Ia's add $10^{51}$~ergs
per SN, while AGB stars add no energy, only metals.  We use \citet{lim05}
yields for Type~II SNe; yields for Type~Ia and AGB stars come from
various works~\citep[see][for details]{opp08}.  Solar metallicity is,
in our case, defined as a total metal mass fraction of 0.0189.

Kinetic outflows are also included emanating from all galaxies.
Gas particles eligible for star formation can be randomly selected to be
in an outflow, by which the particle's velocity is augmented by $v_w$ in a
direction given by {\bf v}$\times${\bf a}, where {\bf v} and {\bf a} are
the instantaneous velocity and acceleration.  The ratio of probabilities
to be in an outflow relative to that to form into stars is given by
$\eta$, the mass loading factor.  Hydrodynamic forces on wind particles
are ``turned off'' until a particle reaches one-tenth the threshold
density for star formation, or a maximum time of 20~kpc/$v_w$, which
attempts to mock up chimneys where outflows escape that would otherwise
be unresolvable in cosmological simulations of appreciable volume.

The choices for $v_w$ and $\eta$ define the ``wind model".  Here we
use scalings expected for momentum-driven winds~\citep{mur05},
though note that such scalings can be generated by other physical
scenarios~\citep{dal08}.  In this model, $v_w = 3 \sigma \sqrt{f_L-1}$,
and $\eta=\sigma_0/\sigma$, where $\sigma$ is the velocity dispersion of
the host galaxy identified using an on-the-fly galaxy finder.  We estimate
$\sigma$ from the galaxy's stellar mass following \citet{mo98}, as
described in \citet{opp08}.  $f_L$ is randomly chosen to lie between
$1.05-2$, based on observations of local starbursts by \citet{rup05}.
This choice also gives outflow velocities consistent with that seen
in $z\sim 1$~\citep{wei08} and $z\sim 2-3$~\citep{ste04} star-forming
galaxies.  $\sigma_0$ is a free parameter, and is adjusted to broadly
match the cosmic star formation history; we choose $\sigma_0=150$~km/s.
This yields $\eta\sim$unity for $L_*$ galaxies at $z=2$, which is in
agreement with the constraints inferred from $z\sim 2$ galaxies by
\citet{erb08}.

This momentum-driven wind model has displayed repeated and often unique
success at matching a range of cosmic star formation and enrichment
data.  This includes observations of IGM enrichment as seen in
$z\sim 2-4$ \ion{C}{iv} systems~\citep{opp06,opp08}, $z\sim 0$
\ion{O}{vi} systems~\citep{opp09} and $z\sim 6$ metal-line
absorbers~\citep{opp09b}.  Concurrently, it also matches the galaxy
mass-metallicity relation~\citep{dav07,fin08}, and the enrichment
levels seen in $z=0$ intragroup gas~\citep{dav08}.  It also suppresses
star formation in agreement with high-redshift luminosity
functions~\citep{dav06,mar07} and the cosmic evolution of UV
luminosity at $z\sim 4-7$~\citep{bou07}.  Hence although this model
is heuristic and does not describe the detailed physics of wind
driving, it appears to regulate star formation in high-$z$ galaxies
in broad accord with observations.

The primary simulation analysed here contains $512^3$ gas and $512^3$ dark
matter particles, in a random cubic periodic volume of 96~Mpc/h (comoving)
on a side with a gravitational softening length (i.e. spatial
resolution) of 3.75~kpc/h (comoving, Plummer equivalent).  We assume
a WMAP-5 concordant cosmology~\citep{kom08}, specifically
$\Omega_m=0.28$, $\Omega_\Lambda=0.72$, $H_0=70$~km/s/Mpc,
$\Omega_b=0.046$, $n=0.96$, and $\sigma_8=0.82$.  This yields gas
and dark matter particle masses of $1.2\times 10^8 M_\odot$ and
$6.1\times 10^8 M_\odot$, respectively.  The initial conditions are
generated with an \citet{eis99} power spectrum at $z=199$, in the
linear regime, and evolved to $z=0$.  We have found through resolution
convergence tests that this is the coarsest possible mass resolution
that adequately models the star formation histories of $z\sim 2-4$
galaxies.  Given that this quarter billion particle simulation took several
months on $\approx 100$ Harpertown cores, it represents the state
of the art for studying rare large galaxies within a random
cosmologically-representative volume.  Fortunately, its volume is
well-matched to current surveys of SMGs.

We identify galaxies using Spline Kernel Interpolative DENMAX (SKID),
and halos using a spherical overdensity criterion; see \citet{ker05}
for more details.  A galaxy's instantaneous star formation rate is the
sum of star formation rates of all star-forming particles in the galaxy.
When we quote metallicities, they are star formation rate-weighted,
since this corresponds most closely to nebular emission line measures of
gas-phase metallicities.  Resolution tests indicate that galaxies with
$\geq128$ star particles have well-converged star formation histories;
for our 96 Mpc/h box, this corresponds to $7.7\times 10^9 M_\odot$.
We take this as our galaxy mass resolution limit.  Note that stellar
mass functions are converged to at least half that mass~\citep{fin06},
but we are being conservative here.  In any case, this paper will be
concerned with substantially more massive systems.  There are 6437
galaxies at $z=2$ in our resolved galaxy sample.

\section{Simulated SMGs}\label{sec:highsfr}

The defining characteristic of SMGs is their high bolometric luminosity,
which most likely implies a high star formation rate.  Hence to identify
SMGs in our models, we make the ansatz that SMGs are the most rapidly
star-forming galaxies in our simulated universe.  In that case, we can
identify simulated SMGs as all galaxies above a chosen SFR threshold
whose number density matches the observed number density of SMGs.
At $z\sim 2$, the observed observed number density is $1-2\times
10^{-5}$~Mpc$^{-3}$~\citep{cha05,tac08}.  Adopting a value of $1.5\times
10^{-5}$~Mpc$^{-3}$, the required threshold for star formation rate in
our simulations is $180\; M_\odot$/yr.  None of our conclusions are
significantly altered had we chosen a SFR threshold higher or lower
within the observationally-allowed range.  Our chosen threshold yields 41
galaxies in our simulated volume of $2.58\times 10^6$~Mpc$^3$.  We will
call this our simulated SMG sample.

\subsection{Stellar mass function}\label{sec:massfcn}

\begin{figure}
\vskip -0.3in
\setlength{\epsfxsize}{0.6\textwidth}
\centerline{\epsfbox{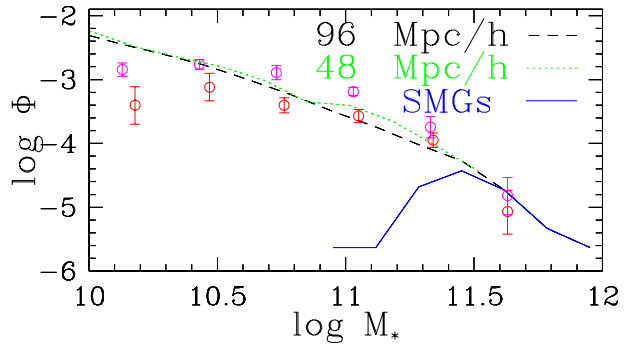}}
\vskip -3.2in
\caption{The stellar mass function (in number per dex per Mpc$^3$) at
$z=2$ from simulations (lines) and observations of \citet[points]{mar09}.
The dashed line shows results from our fiducial 96~Mpc/$h$ volume, while the
dotted green line shows a higher-resolution 48~Mpc/$h$ volume; the
overlap demonstrates good numerical convergence.  The blue line
shows the mass function of our simulated SMG sample.  Data points are
shown at $2<z<3$ (red) and $1.3<z<2$ (magenta).  Agreement with
the simulations is good, particularly for $M>10^{11} M_\odot$,
which is key for this study.  Note that the quoted observational errors 
do not include systematic uncertainties in the redshift
determination or SED fitting; see \citet{mar09} for those.
}
\label{fig:massfcn}
\end{figure}

In Figure~\ref{fig:massfcn} we show the $z=2$ stellar mass function
of our fiducial simulation (dashed line) compared to recent data
from \citet{mar09} at $2<z<3$ (red points) and $1.3<z<2$ (magenta).
The agreement is reasonably good, particularly at the massive end that
is relevant for this work.  This is critical because the number density
of large galaxies (which is a key barometer of SMG models) depends
sensitively on feedback prescriptions, cosmology, etc.  We note that
the models used in \citet{far01} and \citet{fin06} overpredicted the
number densities at the bright end.  We suspect that the calculation of
\citet{dek09}, in which most of the gas accreted at the virial radius
ends up forming stars, would also show this flaw.\footnote{In addition,
the simulation volume of \citet{dek09} is seven times smaller than
that used here, so it cannot statistically model a population of galaxies
as rare as SMGs.} Hence our current simulation, and particularly its
momentum-driven wind feedback prescription, provides a more plausible
platform to investigate the nature of SMGs.

This figure also demonstrates numerical resolution convergence of
our stellar masses.  The green dotted line shows the stellar mass
function from a higher-resolution simulation identical in every way
to our fiducial simulation (including the same number of particles),
except with half the box size (48~Mpc/h), and half the softening length
(1.875~kpc/h).  The agreement in the overlapping mass range with the
larger-volume simulation is excellent.  Overall this shows that our
simulation produces a robust and observationally consistent set of galaxy
stellar masses at $z=2$.

The simulated SMGs (blue line) occupy the most massive end of the stellar
mass distribution.  They range in mass from $10^{11}-10^{12}M_\odot$,
and above $3\times 10^{11}M_\odot$ nearly every galaxy is identified as
a SMG.  Hence, SMGs are among the most massive galaxies at this epoch in
our simulations.  The fact that the most rapidly star forming galaxies
coincide with the most massive galaxies is our first key result, and
our first prediction for the nature of SMGs.

\subsection{Stellar masses}\label{sec:sfrmstar}

\begin{figure*}
\vskip -1in
\setlength{\epsfxsize}{0.8\textwidth}
\centerline{\epsfbox{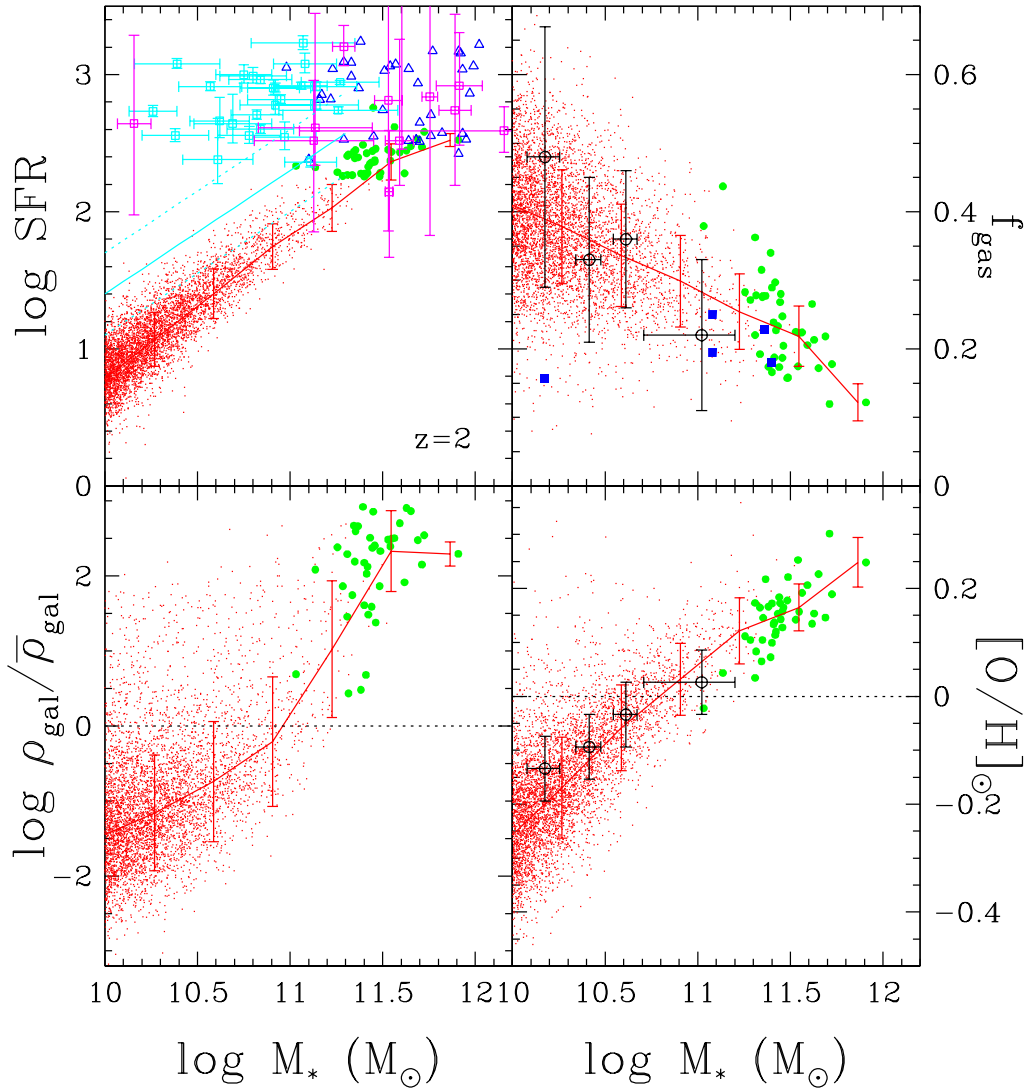}}
\vskip -0.7in
\caption{{\it Top left:} Star formation rate versus stellar mass for the
simulated SMG sample (green circles), all other simulated galaxies (red points),
and observed SMGs from \citet{hai08} (cyan squares), 
\citet{mic09} (blue triangles), and \citet{dye08} (magenta squares).  The
red line shows the median SFR$-M_*$ relation for simulated galaxies,
with a $1\sigma$ scatter shown by the error bars.  The cyan line
shows a fit to SFR$-M_*$ for star-forming BzK galaxies from \citet{dad07},
with approximate $1\sigma$ error bars ($\pm 0.3$~dex) indicated by the dashed lines.
{\it Top right:} Gas fraction $f_{\rm gas}\equiv 
M_{\rm gas}/(M_{\rm gas}+M_*)$ for simulated SMGs (green) and galaxies (red),
plus a running median (red line).  Black circles show
gas fractions inferred for $z\sim 2$ BM/BX galaxies by \citet{erb06}.  
Blue squares show more direct gas fractions from CO measurements by 
\citet{tac08}.
{\it Bottom left:} Local galaxy density smoothed over 1~Mpc spheres.  
SMGs tend to live in denser environments, but particularly at lower
masses there is a substantial spread.
{\it Bottom right:} Oxygen metallicity.  Data points for BM/BX galaxies
are shown from \citet{erb06}.  SMG metallicities tend to be higher than
typical galaxies, following the mass-metallicity relation defined by
lower-mass systems.  Note that the lower-mass SMGs tend to have higher
gas fractions, lower metallicities, and 
(by construction)
higher star formation rates 
than other simulated galaxies of the same $M_*$.
}
\label{fig:sfrmstar}
\end{figure*}

Figure~\ref{fig:sfrmstar} (top left) shows star formation rates versus
stellar mass for our simulated SMG sample (green circles) and for our
resolved galaxy sample (red points).  The simulated SMGs lie on the
SFR$-M_*$ relation defined by the lower mass galaxies, i.e.  these
systems are not outliers as would be expected if their SFRs are being
substantially boosted by mergers.  In fact, there is nothing particular
that distinguishes SMGs from the remainder of the galaxy population,
with the exception of their large stellar masses.

The observational data on SMG stellar masses currently span a wide range,
likely owing to the numerous systematics involved with deriving stellar
masses from photometric near-IR data.  We show a selection of data
from various authors in the figure.  A key point to note is that the
systematic differences in stellar masses between observational samples
are significantly larger than the range of stellar masses within any
given sample.  To illustrate this, one can compare the \citet{hai08}
data (cyan squares) with the \citet{mic09} data (blue triangles).
These two analyses, for the most part, use exactly the same data for
exactly the same galaxies, but employ two different algorithms to obtain
their stellar masses -- yet the \citet{hai08} stellar masses are on
average factor of 6 lower!  In general, other analyses have tended to
find masses agreeing with the higher mass range; the SHADES Lockman Hole
galaxies analysed by \citet{dye08} are shown as the magenta squares,
with SFRs determined from a subset of those for which 350$\mu$m data
was obtained by \citet{cop08} to more accurately characterise the
dust temperature.  \citet{bor05} and \citet{swi06} also find typical
SMG masses well above $10^{11}M_\odot$.  However, it may be that these
analyses have substantially overestimated the stellar masses owing to
contamination in the rest-frame H-band (observed {\it Spitzer}/IRAC)
by AGN~\citep{hai09}.  In contrast, \citet{mic09} argue that the AGN
contamination is small owing to a similar far IR-to-radio flux ratio in
SMGs as that found in local star-forming galaxies.  Clearly, the last
word has not yet been spoken on the stellar masses of SMGs.

Our simulations agree well with the SHADES data and \citet{mic09}
analysis showing high stellar masses.  This is a strong motivating factor
for preferring our interpretation of SMGs.  If instead \citet{hai08}
were correct and the other determinations are biased too high, then our
conclusions would be significantly altered, and our scenario would not
be as viable.  Determining the stellar masses of SMGs to even within a
factor of two would be a major step towards understanding their nature.

\subsection{Star formation rates}\label{sec:sfr}

Turning to the star formation rates, here there is clear disagreement
between the simulations and the observations: the simulated SMGs have
SFRs that are systematically too low by a factor of a few compared to
the real SMGs.  Observationally, SMGs show SFRs from $\sim 400-2000\;
M_\odot/$yr (derived from far-IR data), whereas the simulated SMG range
from $180-570\; M_\odot/$yr.  Hence our star formation rates are too low
by a factor of $\sim 2-4$.  All the observational analyses agree for the
most part, since they typically just translate far-IR flux into SFR
using \citet{ken98a}, though significant uncertainties still remain
particularly owing to poor constraints on the dust temperature.

At face value, the mismatch in SFRs would seem to be a catastrophic
failure of this model; clearly these galaxies need some significant
boost of star formation, such as might occur in a major merger.
Our simulations do not have the numerical resolution to follow the
detailed dynamical processes within merging galaxies that trigger
a starburst, hence one might suppose that they are systematically
missing such a strongly starbursting population.

However, the answer is not so simple.  As we pointed out in the
introduction, hierarchical models predict that the major merger rate
(i.e. $<3:1$) of such massive galaxies is around unity per Hubble
time~\citep{guo08}.  Hence of our 41 simulated galaxies, one would expect
that approximately 1 should be undergoing a major merger, assuming a
$\sim 100$~Myr duration for the SFR boost.  Indeed, we see one galaxy
with a significantly boosted SFR, which upon inspection is undergoing a
merger: It has a SFR of 574~$M_\odot/$yr (the highest among our simulated
SMGs), with a stellar mass of $2.8\times 10^{11} M_\odot$, whereas a more
typical galaxy at this stellar mass has a SFR of $\sim 150 M_\odot/$yr
(we will examine this galaxy further in \S\ref{sec:examples}).  It may be
that had we resolved this galaxy better the starburst might be stronger
and be in better agreement with the observed SMG star formation rates.
But in the larger picture, this cannot explain the boosted SFR seen in
essentially {\it all} SMGs relative to the model expectations.  Hence,
mergers might obtain the correct far-IR fluxes but not the correct number
density at these large masses.

One way to reconcile all of this is if SMGs probed mergers further down
the mass function.  If one out of every 30 galaxies were undergoing a
merger (which is a maximal assumption, since the major merger rate falls
to lower masses), this would imply that SMGs would have to have masses
down to $\sim 3\times 10^{10} M_\odot$ to achieve the correct number
density.  In that case, the observed large stellar masses would be poorly
fit in the models, although this might be better accommodated if the
stellar masses from \citet{hai08} are correct.  This is (qualitatively)
the dilemma faced by the semi-analytic models of \citet{bau05}, which as
discussed in~\citet{swi08} match the number density but fail to match the
stellar masses.  Furthermore, such low-mass systems would seem to require
an implausibly top-heavy IMF to produce the SMG far-IR luminosities.

In short, our simulations broadly match the number densities and the
stellar masses of SMGs but fail to reproduce the observed SFRs by a
modest factor.  Other popular scenarios fail to match different ones
of these three quantities.  It is not clear where the answer lies,
but in \S\ref{sec:imf} we will examine how the SFR discrepancy might be
symptomatic of a more general problem with $z\sim 2$ galaxies.  For now
we will focus on examining other properties of our simulated SMGs to see how
our scenario fares against other observational data, and to gain a 
greater physical insight into the nature of this population.

\subsection{Gas fractions}\label{sec:gasfrac}

The upper right panel of Figure~\ref{fig:sfrmstar} shows the gas fractions
($\equiv M_{\rm gas}/(M_{\rm gas}+M_*)$) of our simulated galaxies,
with the subsample of SMGs indicated by the green points.  Binned mean
gas fractions from a sample of BM/BX galaxies (i.e.  $z\approx 2$
photometrically-selected galaxies) are shown from \citet{erb06}.
It is worth cautioning that this comparison is fairly crude; first,
the simulated gas masses are taken to be the mass of star-forming
gas in each galaxy, which may not correspond directly to what would
be inferred from molecular or \ion{H}{i} emission.  Furthermore, the
\citet{erb06} data do not come from direct measurements of gas masses,
but rather are gas masses inferred from the star formation rate surface
density assuming the \citet{ken98} relation.  Despite these caveats,
the simulated galaxy population shows a general agreement with the
trend of higher gas fractions in lower-mass systems.  Note that this is
critically dependent on our outflow model; other outflow models do not
match as well~\citep{dav07}.

Direct measures of gas fractions at $z\sim 2-3$ are now becoming
available thanks to improving millimetre interferometry.  CO gas
measures by \citet{tac08} are shown as blue squares in the figure.
Unfortunately, once again the comparison is not straightforward, since
the observations include only H$_2+$He, while the simulations include
all the cold gas including \ion{H}{i}.  Also, the observed measurements
depend on the famous X factor, the conversion factor from CO emission to
molecular hydrogen mass; the data of \citet{tac08} is best fit by an X
factor similar to that found in local ULIRGs, which is about $5\times$
lower than that in the Milky Way disk.  Still, the data generally lie in
the range of the simulated galaxies.  The one exception is the ``Cosmic
Eye''~\citep{sma07,cop07} at $M_*=1.5\times 10^{10}M_\odot$, which is the
blue point well below the relation at low masses; it could be a $2\sigma$
outlier, or its $M_*$ or $f_{\rm gas}$ could be misestimated, or it may
have a large \ion{H}{i} reservoir.  Even so, there are still simulated
galaxies with similarly low $f_{\rm gas}$ at these stellar masses.

Although simulated SMGs broadly follow the $f_{\rm gas}-M_*$
anti-correlation seen in lower-mass systems, it is notable that at a
given stellar mass, SMGs tend to be more gas-rich objects.  This simply
reflects the fact that SMGs are a SFR-selected sample, which in turn
selects galaxies that have large gas reservoirs to fuel star formation.
While the simulated galaxies are broadly in agreement with the data,
more direct measures of gas mass, together with a better understanding
of the relationship between \ion{H}{i} and molecular gas, can potentially
provide more robust and interesting constraints.

\subsection{Metallicities}\label{sec:metal}

The bottom right panel of Figure~\ref{fig:sfrmstar} shows the metallicity
of simulated galaxies versus $M_*$.  The SMGs lie along a well-defined
mass-metallicity relation that broadly matches an extension of the
relation for $z\sim 2$ BM/BX galaxies~\citep{erb06,fin08}.  Because
they are extremely massive, they also have quite high metallicities,
typically solar or more.  Note that the amplitude agreement is perhaps
not so meaningful given the systematic uncertainties in metallicity
measures of distant galaxies~\citep[e.g.][]{kew08}, but the relative
metallicities (and hence the trend with $M_*$) are likely more robust.

At a given $M_*$, the SMGs tend to have lower metallicities; this
goes along with their gas richness and high star formation rate.
In \citet{fin08} we pointed out that galaxy metallicities are set by
a competition between accretion (which dilutes the metallicity) and
star formation (which increases it), the balance of which is modulated
by outflows.  Hence, a galaxy that has recently acquired a large amount
of gas through an infall event will have a lower metallicity, as well as
a higher gas content and star formation rate.  This predicted trend is
observed in local galaxies~\citep{ell08,pee09}.  It will be interesting
to see, as the data improves, whether this trend is also true in SMGs compared
to similar mass-selected galaxy samples.

\subsection{Environment and clustering}\label{sec:envir}

The bottom left panel of Figure~\ref{fig:sfrmstar} shows the local
galaxy density of our simulated galaxy sample spherically averaged
over 1~comoving~Mpc, plotted versus stellar mass.  The local density is
computed using all the resolved galaxies ($M_*>7.7\times 10^9M_\odot$).
Our simulated SMGs clearly prefer to live in overdense environments.
However, at a given stellar mass there is no particular trend; SMGs and
non-SMGs are not obviously distinguished by environment.  Hence the
dense environs simply reflect the high degree of clustering expected
for a massive galaxy population.

Observational estimates of SMG environments are difficult, owing to
the relatively small and non-uniform coverages of existing surveys.
\citet{ser08} found that SMGs tend to prefer dense environments (though
not exclusively so), although the statistics were only definitive for
$1<z<1.5$ systems.  \citet{tam09} studying the SSA22 protocluster regions
at $z=3.1$ also finds evidence that SMGs prefer to live in denser regions.
\citet{cha09} finds that SMGs appear to be more biased than Lyman
break galaxies.
The idea that SMGs tend to live in dense environments is
qualitatively consistent with our model, but a quantitative comparison
will have to await improved data sets, particularly over wider areas;
SCUBA-2 will help in this effort.

\begin{figure}
\setlength{\epsfxsize}{0.5\textwidth}
\centerline{\epsfbox{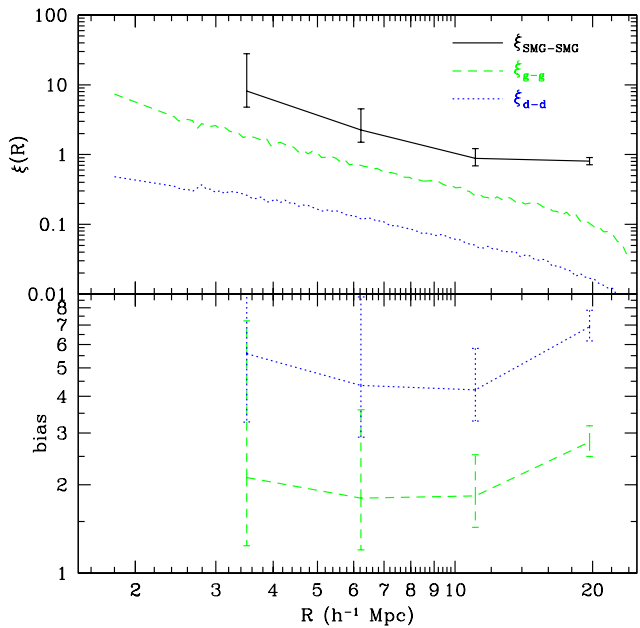}}
\caption{{\it Top:} Correlation function of our simulated SMGs (black
solid), of our resolved galaxy sample (green dashed), and the dark matter
distribution (blue dotted).  Error bars on the SMG curve show Poisson
errors in the pair counts; the Poisson errors for the other curves are
negligibly small and are likely dominated by cosmic variance, which is
not included here.
{\it Bottom:} The bias of SMGs estimated from $\sqrt{\xi_{\rm SMG}/\xi}$
relative to the dark matter (blue dotted), and of all galaxies (green dashed).
SMGs are a highly clustered population with $r_0\sim 10\hmpc$ and
a bias of $\sim 6$.
}
\label{fig:corr}
\end{figure}

Our large simulation volume allows us to make predictions for the
clustering of SMGs.  Figure~\ref{fig:corr} (top panel) shows the
autocorrelation functions of simulated SMGs (solid black) at $z=2$,
along with that of our entire resolved galaxy sample (having a
number density of $2.5\times 10^{-3}$~Mpc$^{-3}$, similar to that of bright
Lyman break galaxies) and the dark matter.  The autocorrelation of
SMGs clearly shows a larger amplitude than the full galaxy sample,
while its slope is fairly similar.  The clustering length of simulated
SMGs is $r_0\approx 10 \hmpc$, compared to $5.2 \hmpc$ for the
full resolved galaxy sample.

In the bottom panel we show the bias of SMGs relative to all resolved
galaxies (green) and to the dark matter (blue).  We estimate the bias by taking
the square root of the ratio of the correlation functions.  Relative to the
dark matter, the bias of SMGs is $\approx 6$, which is about twice that of
typical galaxies.  In short, SMGs are highly biased and clustered objects
in our simulation, as would be expected from their large stellar masses.

Observations of SMGs support the idea that they are highly clustered.
\citet{bla05} found a correlation length of $r_0\approx 7.5\pm 2.6
\hmpc$ for SMGs, suggesting that they are located in fairly massive
halos.  \citet{far06} determined $r_0\approx 6\pm 1\hmpc$ for ULIRGs
at $1.5<z<3$, though many of these had IR luminosities below that
of SMGs.  These determinations are somewhat lower than we predict,
although given the uncertainties, the agreement is not bad.

The general prediction that simulated SMGs live in fairly dense and
clustered environments seems to be borne out, at least preliminarily,
in available data.  But a detailed comparison will have to await
large uniform samples of SMGs as is anticipated from upcoming
instruments.  Clustering may provide a key discriminant of SMG
scenarios, since if SMGs are mergers of more ordinary-sized galaxies,
they will generally cluster as such, although a weak effect is
expected from merger bias \citep[i.e. that more massive galaxies
have a higher merger rate; e.g.][]{guo08,cha09}.  If SMGs are
massive, they have little choice but to live in massive halos that
cluster strongly.

\section{Example Simulated SMGs}\label{sec:examples}

\begin{figure}
\vskip -0.3in
\setlength{\epsfxsize}{0.6\textwidth}
\centerline{\epsfbox{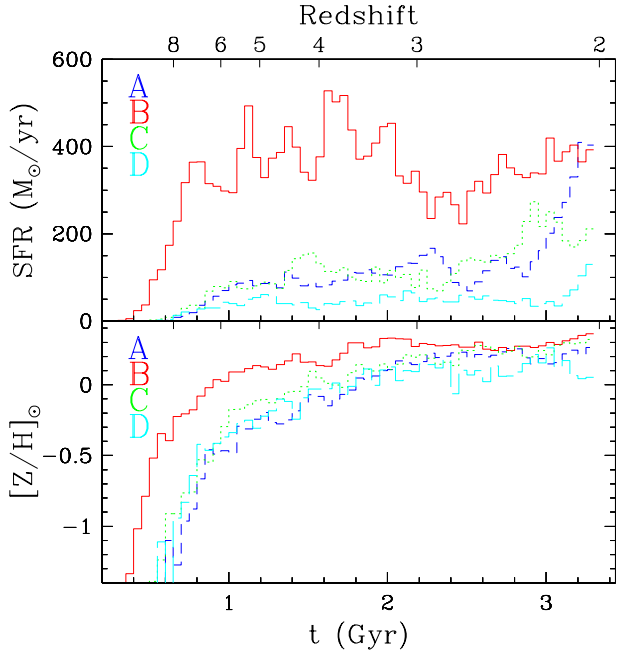}}
\vskip -1.8in
\caption{{\it Top:} Star formation histories in 50~Myr bins for the four 
galaxies described in text.
The right edge of the histograms corresponds to $z=2$, 
the adopted epoch of observation.
{\it Bottom:} Mean metallicity of star formation as a function of time
for the four chosen galaxies.
}
\label{fig:sfhist}
\end{figure}

\begin{figure*}
\setlength{\epsfxsize}{1.0\textwidth}
\centerline{\epsfbox{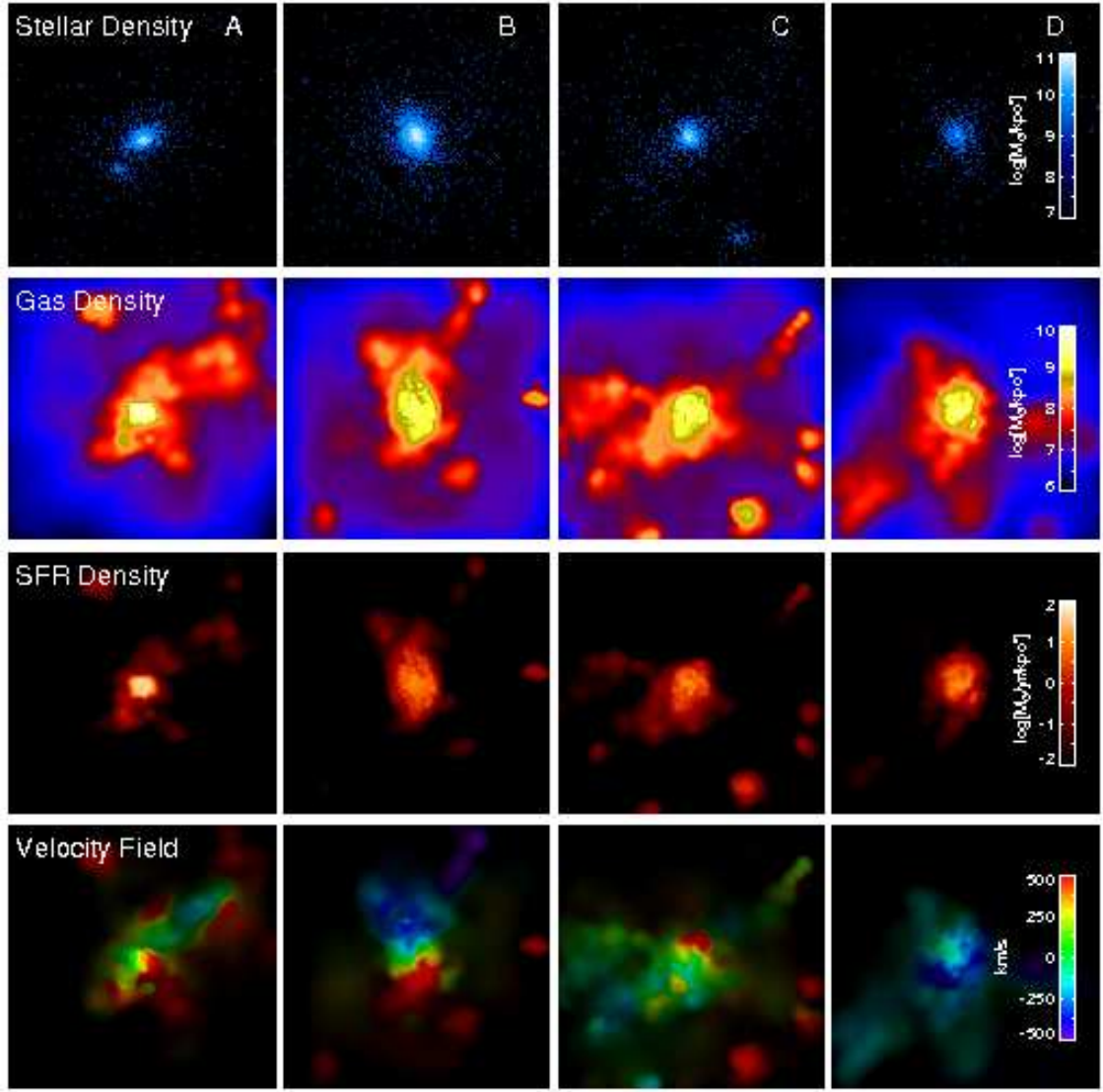}}
\caption{Simulated $z=2$ SMGs A, B, C, D (left to right), in slices 50~kpc
(physical; 6") across, showing from top to bottom the stellar density,
gas density, star formation rate density, and line-of-sight velocity
field.
}
\label{fig:4gals}
\end{figure*}

We now concentrate on several individual simulated SMGs to study their
properties in more detail.  We choose four galaxies that span the range
of SMGs from our $z=2$ simulation output:
\begin{itemize}
\item[{\bf A.}] The highest SFR SMG ($M_*=2.8\times 10^{11} M_\odot$, SFR=$574 M_\odot/$yr);\\ %54752
\item[{\bf B.}] The most massive SMG ($M_*=7.9\times 10^{11}$, SFR=$334 M_\odot/$yr);\\ %56596
\item[{\bf C.}] The median $M_*$ SMG ($M_*=2.7\times 10^{11}$, SFR=$224 M_\odot/$yr);\\ %73329
\item[{\bf D.}] The least massive SMG ($M_*=1.1\times 10^{11}$, SFR=$216 M_\odot/$yr). %82821
\end{itemize}
We also refer the reader to our website,
{\tt http://luca.as.arizona.edu/$\sim$oppen/IGM/submm.html},
where we show images similar to Figure~\ref{fig:4gals} 
for all of our simulated SMGs.

\subsection{Star formation and enrichment histories}\label{sec:sfh}

Figure~\ref{fig:sfhist}, top panel, shows the star formation histories
(SFHs) of these four galaxies.  Overall, the SFHs are relatively constant
since $z\sim 6$ ($t=1$~Gyr), with the larger stellar masses simply
owing to a higher rate of quiescent SF.  Excursions at the $\sim\times2$
level from the time-averaged SFR are not uncommon.

However, in their recent histories, i.e. in their last $\sim 200$~Myr,
our SMGs divide into two types, ones whose current SFR is comparable
to their recent average (B,C) and ones that are experiencing a current
boost in their SFR (A,D).  This is directly related to where they lie
in the SFR$-M_*$ plane (Figure~\ref{fig:sfrmstar}): A and D are clearly
outliers above the median SFR, while B and C lie close to the median.
Hence, we can generalise these trends by looking at where SMGs lie
relative to other galaxies in the SFR$-M_*$ plane.

The careful reader will note that the final $z=2$ SFRs in
Figure~\ref{fig:sfhist} are not always identical to the SFRs listed for
these objects in \S\ref{sec:examples}.  The difference is that the plot
shows 50 Myr-averaged star formation rates, while the numbers quoted
above are instantaneous SFRs.  For the more quiescent systems, B \& C,
the numbers are similar but for the bursting systems, A \& D, the
instantaneous rate is non-trivially higher than the past 50 Myr smoothed
average.

While the scatter in SFR$-M_*$ is fairly small ($0.3$~dex), this is
still comparable to the spread in SFRs for SMGs, which means that a
SFR-selected sample like SMGs will preferentially pick out galaxies like
A and D that are undergoing a boost.  Therefore, despite an underlying trend
in SFR$-M_*$, examining only SMGs in current samples would likely result
in a scatter plot in this plane, as shown in Figure~\ref{fig:sfrmstar}.
As one probes to lower masses, SMGs more strongly select high-SFR outliers
in SFR$-M_*$.

Figure~\ref{fig:sfhist}, bottom panel, shows the growth of metallicity
in these galaxies.  All four galaxies enrich themselves to at least
50\% solar by $z=6$, showing that even with strong outflows that drive
the majority of metals into the IGM, massive galaxies still enrich
themselves quite rapidly.  The most massive galaxy has the earliest
formation epoch and hence the earliest metal growth, but by $z\sim 4$
all the galaxies have similar metallicities around solar.  A consequence
of rapid enrichment is that there is also expected to be large amounts
of dust in these galaxies from a fairly early epoch (assuming that dust
production tracks metallicity on a reasonably short timescale), providing
the obscuration required to produce the large far-IR luminosities.
We leave for future work an analysis of dust reprocessing in these
systems to predict a detailed SED~\citep[as in][]{nar09a}; this will
likely require substantially higher spatial resolution than we can
currently achieve in representative cosmological volumes.

\subsection{Morphologies and kinematics}\label{sec:morph}

One argument in favour of a merger origin for SMGs is their apparently
compact and disturbed morphologies.  Double nuclei, tidal tails, and
kinematics that depart strongly from simple spiral or elliptical galaxies
are seen in many cases~\citep[e.g.][]{tac06,tac08}.  How can a model
that postulates that these objects are oversized but otherwise normal
star-forming galaxies be reconciled with such a scenario?  The short
answer is, at this epoch {\it most} star-forming galaxies are compact and
disturbed~\citep{gia02}, hence such traits are not necessarily indicative
of an ongoing merger as would be the case locally~\citep{law07}.

Figure~\ref{fig:4gals} shows $50\times 50$~kpc (physical; about 12"
at $z=2$) maps of our four example simulated SMGs, in various physical
quantities.  Galaxy A (left panels) is undergoing a merger at this
epoch with a smaller object about one-eighth its mass, seen just to the
lower left.  This results in a more centrally-concentrated distribution
of gas and stars compared to the other systems, as well as a fairly
chaotic velocity field.  We note that the smaller object was originally
significantly larger, about twice its current mass, but was tidally
disrupted to its current size during the merger.

The largest galaxy, B, appears the most quiescent and ordered of them all.
It shows an extended gaseous star-forming centre with a large central
galaxy showing ordered rotation.  It is being fed by gas along various
streams, but while some lumps are apparent, the majority 
comes in relatively smoothly.  As argued in \citet{ker08}, that is
typically the case for high-$z$ galaxies.

The other two galaxies (C,D) show more typical examples of simulated SMGs.
In both cases they show relatively extended gas and star-forming regions,
at least compared to A.  However, the velocity fields do not show any
particular order, as there are a fair number of other small galaxies
in the vicinity as well as infalling gas resulting in somewhat chaotic
kinematics.

The actual appearance of these galaxies in observable bands depends on a
number of other effects.  CO traces neutral gas density to some extent,
modulo variations in the molecular gas fraction and the ratio of molecular
gas to CO.  Ongoing star formation can be traced directly in the far-IR
continuum, although the relationship of the reradiating dust to the star
formation may not be trivial.  Radio observations may provide a more
direct constraint, assuming that any contribution from a central AGN can
be resolved out.  The stellar density can be most directly traced in the
rest near-IR, but given the amount of dust in SMGs, even those wavelengths
can be substantially affected by extinction and its associated patchiness.
The velocity field traced in H$\alpha$ with integral field units such
as SINFONI~\citep{for09} provides a fairly direct measure of the star
forming gas kinematics, if not extincted.  All these observations are
possible today, but their interpretation is complex.  Dust and molecular
radiative transfer models as in \citet{nar09b} are required for a proper
interpretation; in a fully cosmological setting this is a strong challenge
for numericists in the coming years.

Using millimetre interferometry plus other wavelength data, \citet{tac06}
provide a variety of evidence that SMGs are scaled-up versions of
local major mergers seen as ULIRGs.  Many of the arguments involve
the compactness of the star formation as seen in CO lines, typically
$1-2$~kpc, although these were high-CO transitions that typically come
from only the densest gas.  On the other hand, there is good evidence
from radio interferometry that the star formation in SMGs is extended
over many kpc scales (as in our simulated SMGs in Figure~\ref{fig:4gals})
and not confined to the nucleus as in local ULIRGs~\citep{cha04,big08}.
Also, SMGs appear significantly less reddened than nuclear starbursts,
which further differentiates them from nearby ULIRGs~\citep{hai09}.
This may argue against the major merger interpretation, since in mergers
the star-forming gas is generally driven into a very compact and
highly obscured nuclear
starburst that in local ULIRGs are $\la 1$~kpc, although simulations of
SMGs from mergers do show larger extents of a few kpc~\citep[e.g.][]{nar09b}.
Hence, current results on CO morphologies are not conclusive at to whether
or not SMGs are major mergers or not, and it remains possible that SMGs
could be a heterogeneous population.  The kinematics are also not yet
conclusive, as even our non-merging systems show disordered kinematics.
A key advance in this area would be high-resolution imaging in the
near-IR, with, e.g., {\it Hubble}/WFC3 or JWST, to probe the stellar
distribution on sub-arcsec scales (modulo patchy extinction).  We predict
single smooth stellar distributions in most cases with perhaps small
companions, whereas a pre-coalescence major merger would clearly show
two distinct large concentrations.

\section{Relationship to other galaxy populations}\label{sec:imf}

\subsection{Are the observed star formation rates overestimated?}

The overall properties for SMGs derived from simulations broadly
agree with the available observations of SMGs, including their stellar
masses, gas fractions, morphologies, and clustering.  However, there
is still one clearly discrepant issue: the star formation rates in
the simulated galaxies are a factor of a few below that inferred
from far-IR measures of SMG SFRs.  Is this discrepancy sufficiently
serious to rule out our scenario for SMGs, or is there a plausible
explanation?

\begin{figure}
\vskip -0.3in
\setlength{\epsfxsize}{0.6\textwidth}
\centerline{\epsfbox{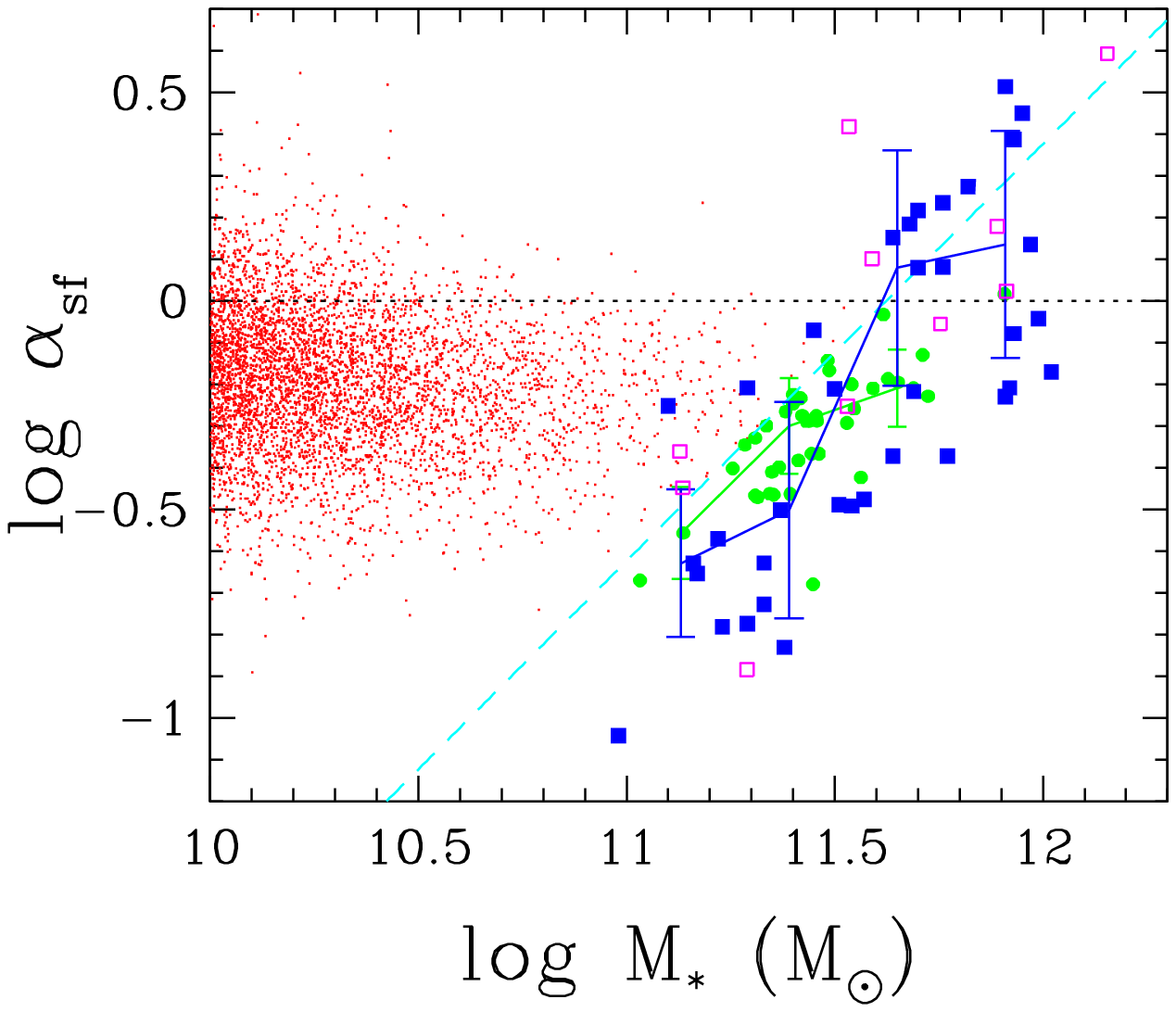}}
\vskip -2.8in
\caption{The star formation activity parameter $\asf\equiv (M_*/{\rm
SFR})/(t_H-1\;{\rm Gyr})$ for the simulated SMGs (green) and all the simulated
galaxies (red), with green solid lines showing the running median for the SMGs.
The cyan dashed line shows a limit of SFR$=180\; M_\odot/$yr
at $z=2$.  Observations of SMGs from \citet{mic09} and \citet{dye08}
with the SFRs arbitrarily lowered (and thus $\asf$ raised) by a factor of 3
are plotted as blue triangles and magenta squares, respectively, with a
running median shown for the \citet{mic09} sample.  Such a
rescaling places the median value of $\asf$ in line with simulated SMGs.
}
\label{fig:sfact}
\end{figure}

To investigate this, we introduce the star formation activity parameter
\begin{equation}
\asf\equiv (M_*/{\rm SFR})/(t_H-1 {\rm Gyr})
\end{equation}
\citep{dav08b}.  $\asf$ measures the fraction of a Hubble time (minus
the first Gyr in which little growth happens) that this galaxy must form
stars at its current rate to produce its current stellar mass.

Figure~\ref{fig:sfact} shows $\asf$ for the simulated SMGs (green points),
which indicates that $\asf\approx 0.8$ for simulated galaxies in this
mass range.  Taking the data of \citet{dye08} and \citet{mic09} at
face value yields a typical $\asf\approx 0.2-0.3$.  If we apply an ad
hoc correction factor of $\times 3$ to their $\asf$, then the results
are as shown by the blue and magenta data points.  This results in
good agreement between the simulated and observed SMGs, with a median
$\asf\approx 0.7-0.8$.  This shows that to reconcile our model with data,
$M_*/$SFR must be overestimated in the models or underestimated in the
data by about a factor of three.

Let us first consider whether the simulations might underestimate
the star formation rates by $\sim\times 3$, perhaps owing to some
details of the subgrid star formation prescription.  It turns out
this is not viable in a cold mode accretion-driven scenario, where
the star formation rate is strongly limited by the gas supply rate,
which in turn is set predominantly by the gravitational accretion
rate~\citep{ker05,dek06,ker08}.  That is the reason why $\asf\sim 1$
is a fairly generic result attained in both numerical and semi-analytic
galaxy formation models~\citep{dav08b}.  Hence, the model galaxies cannot
form stars faster unless gravity brings gas in faster.  Alternatively,
we argued in \S\ref{sec:sfr} that mergers are unlikely to be responsible
for such an underestimate, since only a small fraction of galaxies are
undergoing a merger at any given time.

One could equivalently match $\asf$ by claiming that the observed
stellar masses are underestimated by $\sim\times 3$.  However, this
would then require galaxies with quite large stellar masses, which
grossly exacerbates the problem of producing enough such systems
by $z\sim 2$ in a hierarchical cosmology.  Note that the recently
discussed stellar population issue of thermally pulsating AGB
stars~\citep{mar05} would tend to lower, not raise, the inferred
stellar masses, though it is not expected to be a large factor
compared to the current systematic uncertainties.

Hence for our scenario to be viable, the observed star formation
rates in SMGs must have been overestimated by a (modest) factor of
3.  It seems rather presumptuous to claim based on our models that
observations are wrong, but a factor of 3 does not seem out of the
realm of possibility, given the current uncertainties involved in
converting various fluxes to star formation rates in high-$z$
galaxies.

As an aside, a subtle predicted trend is that $\asf$ is correlated
with $M_*$.  This is characteristic of a SFR-selected sample; the
dashed cyan line in Figure~\ref{fig:sfact} shows our adopted SFR limit
of 180~$M_\odot/$yr.  Smaller SMGs would be preferentially selected to
have elevated SFRs (and hence lower $\asf$), while more massive galaxies
are more quiescent and so would have $\asf\sim 1$.  This trend is also
seen in the observed samples but only weakly, probably owing to the
currently large systematic uncertainties in determining $M_*$ and SFR.

There are other lines of evidence that suggest that SFRs in $z\sim 2$
galaxies of {\it all} types have been overestimated.  As pointed out in
\citet{dav08b}, this would reconcile the observed values of $\asf$ in BzK
galaxies~\citep{dad07} with the simulated $\asf$.  More such evidence
comes from examining the global cosmic star formation rate density
evolution versus stellar mass growth~\citep{wil08}; they showed that
the time derivative of the cosmic stellar mass density at $z\sim 2$ is
lower by $\sim\times 3$ compared to the star formation rate inferred from
high-mass tracers such as H$\alpha$ or UV~\citep[though see][]{red09},
suggesting that the true SFR is lower than that obtained using conversion
factors from \citet{ken98a}.

Possible physical reasons for such an overestimate are numerous.
It could simply be a calibration issue; the locally-calibrated conversions
between far-IR, UV or radio luminosities could be off by this factor in
the case of high-$z$ galaxies.  Galaxies as active as SMGs do not exist
in the local Universe, so it is difficult to test such calibrations in
analogous systems nearby.  One could imagine that the physical conditions
in the ISM are systematically different, since high-$z$ galaxies tend
to have much higher star formation rate surface densities than typical
galaxies today ~\citep{erb06,tac08}.  It could also be an issue with dust
temperatures, as the inferred SFR is highly sensitive to its assumed
value \citep[e.g.][]{far01}.  Observations of 350$\mu$m and 850$\mu$m
fluxes by \citet{cop08} indicate a typical dust temperature of 28~K
(with substantial scatter), which is lower than the canonically-assumed
35~K from local starbursts~\citep{dun00}; this will be better determined
with upcoming {\it Herschel} data.  The lower temperature could imply
as much as $\times 2$ lower SFRs.  There also may be some far-IR light
from AGN, particularly for the most luminous sources, although direct
estimates show that it is likely to be sub-dominant in SMGs.  Finally,
SED fitting itself can be uncertain at a factor of $\sim 2-3$ level
in SFR when only broad-band data are employed~\citep{muz09}, although
it's not clear that this would provide a systematic shift.  It is not
a stretch to think that one or more of these factors could lead to a
factor of 3 overestimate in the SFRs.

Another perhaps related possibility is that the IMF could be different
in high-$z$ galaxies.  This has been offered as a solution to various
nagging quandaries in galaxy and stellar evolution, such as the
colour and luminosity evolution of early-type galaxies~\citep{van08},
the abundance of carbon enhanced metal-poor stars in the Milky
Way~\citep{tum07}, and the evolution of the cosmic stellar mass
assembly~\citep{far07,wil08}.  One solution invoked to solve all these
problems, as pointed out in \citet{dav08b}, would be that the IMF
is somewhat more top-heavy or bottom-light at higher redshifts,
such that SFRs would be overestimated by $\sim\times 3$ at $z\sim
2$.  Recently, \citet{meu09} suggested that nearby galaxies with
higher SFR surface densities have more top-heavy IMFs, which adds
some local credibility (albeit controversial) to the idea of the
IMF perhaps being more top-heavy (or bottom-light) at early epochs.

The level of IMF variation required in our scenario is far milder than
that in the \citet{bau05} semi-analytic models of SMGs, and it is not
easy to constrain directly even in local galaxies let alone at $z\sim 2$.
It is, for instance, quite consistent with direct constraints on the
IMF of SMGs by \citet{tac08}.  IMF variations would also have other
implications, such as more metals (and hence dust) being produced
for a given stellar mass, which could have important effects on dust
temperatures and inferred SFRs~\citep[see e.g.][]{bau05}; a thorough
analysis of all its effects is beyond the scope of this work.  It must
be pointed out that there is no clear evidence for IMF variations in any
context, so this possibility should be considered with some skepticism.

For our scenario, it is irrelevant whether the cause of the purported
SFR overestimates is a different IMF or some calibration effect.
The main point is that if the SFRs have been systematically
overestimated by a factor of 3, then this removes the main clearly
discrepant aspect of our model when compared to current observations
of SMGs.  Such an overestimate may not be unique to SMGs, but may
be reflective of a systematic difference in all $z\sim 2$ star-forming
galaxies relative to the local ones.

\subsection{Duration of the SMG phase}\label{sec:dutycycle}

SMGs in our scenario are forming stars relatively quiescently, as seen in
Figure~\ref{fig:sfhist}.  Hence one expects that SMGs remain identifiable
as SMGs for a long time.  This contrasts with the short duty cycle of
SMGs in a merger-based scenario.  Inferred duty cycles for SMGs tend
to be short, but these are mostly based on gas consumption timescales.
In our scenario, such timescales are not relevant because the IGM supplies
gas at a rate comparable to the star formation rate.  Therefore, our
SMGs don't ``use up" their gas but merely consume it at the rate at which
it is being supplied.  The key point is that simulations generically
predict remarkably large gas supply rates in high-$z$ galaxies.

\begin{figure}
\vskip -0.5in
\setlength{\epsfxsize}{0.55\textwidth}
\centerline{\epsfbox{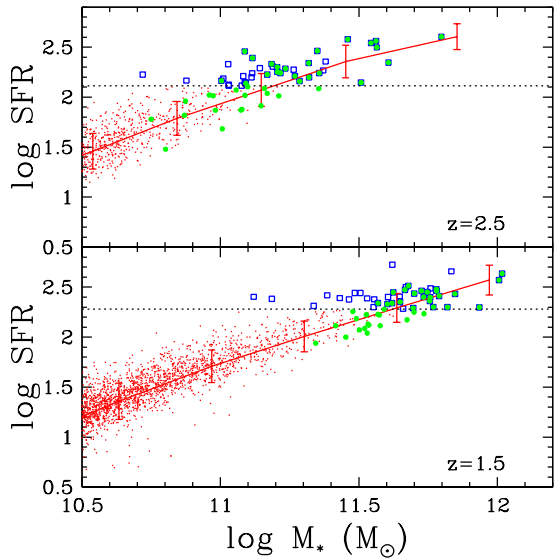}}
\vskip -1.7in
\caption{SFR vs. $M_*$ at $z=2.5$ (top panel) and 
$z=1.5$ (bottom panel).  Descendants/progenitors of $z=2$ simulated SMGs
are shown as green points.  Blue squares denote SMGs that would be
identified based on a SFR threshold at those redshifts; the SFR thresholds
are indicated by the dotted lines.  More than half
the $z=2$ SMGs are identified with similarly-selected SMGs at $z=2.5$ 
and $z=1.5$.
}
\label{fig:progen}
\end{figure}

To get a handle on the duration of the SMG phase, we identify
the progenitors and descendants of $z=2$ SMGs in our simulations
at $z=2.5$ and $z=1.5$, i.e. 0.67~Gyr prior and 1~Gyr later.
In Figure~\ref{fig:progen} we show the location in SFR$-M_*$ of $z=2$
SMGs (green points) at these earlier and later epochs.  For reference,
the horizontal line demarcates the SFR threshold needed to maintain a
number density of $1.5\times 10^{-5}$~Mpc$^{-3}$; the values are slightly
different than at $z=2$, namely $130\; M_\odot/$yr at $z=2.5$ and $190\;
M_\odot/$yr at $z=1.5$.

At both $z=2.5$ and $z=1.5$, slightly more than half the $z=2$ SMGs
are among the galaxies that would be identified as SMGs based on an
analogous SFR threshold.  The remainder fall just below the SFR threshold.
In general, the $z=2$ SMGs at these other epochs are fairly randomly
distributed in SFR amongst the general galaxy population at a given $M_*$.
This is in line with our scenario that SMGs have a fairly constant SFR,
with minor ($\sim\times 2$) excursions owing to stochastic accretion
events.

Expressed 
in terms of a duty cycle, one might say that the SMGs in our
simulations have a duty cycle of $\sim50$\%, since at any given epoch the
high-SFR criterion picks out about half the massive galaxy population.
This duty cycle depends on mass; the lower-mass SMGs tend to be greater
outliers, and hence have smaller duty cycles.  Of course, the concept of
a duty cycle may not be exactly appropriate for these systems, given that
they do not become dramatically less active when they are not in the SMG
phase, but rather simply fall slightly below what would be identified
in an SFR-selected sample.

One implication of this relatively high duty cycle is that the
number density and clustering of SMGs should independently translate
into roughly the same halo mass.  This would indicate that a large
fraction of halos at the high-mass end are occupied by SMGs.  This
is in contrast to a scenario where SMGs are major mergers, in which
the small duty cycle of SMGs in a merger phase would imply a low
fraction of massive halos containing an SMG.  With wider and more
uniform samples, this prediction can be tested relatively
straightforwardly, which we expect will strongly discriminate between the 
different SMG scenarios.

\subsection{SMG descendants today}\label{sec:descendants}

Where do the descendants of SMGs live today?  In our simulation we
can follow our simulated SMG sample to $z=0$, where we can directly
identify the halos in which they reside.

\begin{figure}
\vskip -0.4in
\setlength{\epsfxsize}{0.65\textwidth}
\centerline{\epsfbox{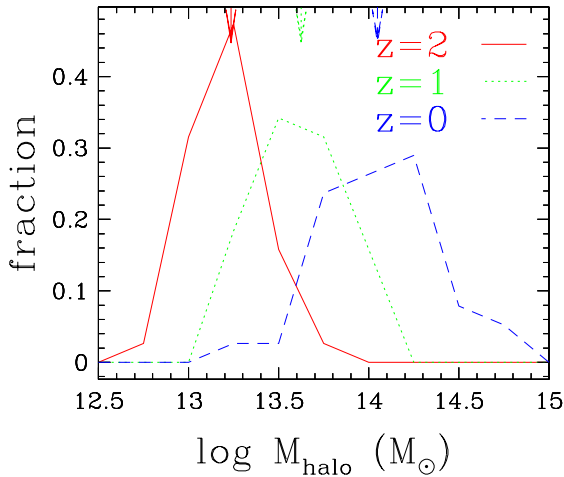}}
\vskip -3.1in
\caption{Histogram of halo masses of simulated SMGs 
at $z=2$ (red solid line), and the halos in which their
descendants live at $z=1$ (green dotted) and $z=0$ (blue dashed).
The median values at each epoch are indicated at the top of the figure.
}
\label{fig:haloevol}
\end{figure}

Figure~\ref{fig:haloevol} shows a histogram of the halo masses of
simulated SMGs at $z=2$ (red solid line), and the halos in which
their descendants live at $z=1$ (green dotted) and $z=0$ (blue
dashed).  At each epoch, these are among the most massive halos in
the simulation.  From $z=2\rightarrow 0$, the typical halo mass
grows from $1.7\times 10^{13} M_\odot$ to $1.1\times 10^{14} M_\odot$.
The latter is a group-sized halo.  Hence, we can infer that SMGs are
typically the brightest group galaxies in the process of active formation
at $z\sim 2$.  This is also consistent with their number density and large
clustering length.

Another issue is that SMGs cannot continue to form stars at their
$z\sim 2$ rates all the way to $z=0$, since no galaxies approaching
those bolometric luminosities are seen locally, and their stellar
masses would be far too large.  Our simulated SMGs match the $z\sim
2$ observed stellar mass function (Figure~\ref{fig:massfcn}).  How
much longer can they sustain these rates and still be consistent
with lower-$z$ mass functions?

To answer this, let us make the ansatz that SMGs turn into the most
massive galaxies today.  Clearly this is not strictly true, but
given that the simulated SMGs are among the most massive galaxies at $z\sim 2$
living in the most massive halos, it may not be a bad approximation.
In any case it provides a upper limit to how much SMGs can grow.
Examining the stellar mass functions versus redshift of \citet{mar09},
it is clear that there is not much room for growth at the massive
end.  For SMG number densities of $10^{-5}$~Mpc$^{-3}$, the stellar
mass has grown by only $0.2$~dex since $z\sim 2$.  Hence even if
one were to ignore dry merging, this only allows for a growth of 60\%
in stellar mass in the last 10~Gyr.  If the typical stellar mass
is $M_*=10^{11.3} M_\odot$, then this is a growth of $\approx
1.2\times 10^{11}M_\odot$.  Taking a median SFR of $\approx 400
M_\odot/$yr~\citep{cop08}, this means a typical SMG can only continue
forming stars at this rate for at most 300~Myr, and perhaps much
less given that massive galaxies also grow by dry
merging.  Therefore, one would be observing SMGs at a fairly
special time, when the star formation has to shut off within $\ll 0.1t_H$. 
If, however,
the SFRs were lowered by a factor of three, the SMGs could
conceivably continue to form stars at their observed rates for the
better part of a Gyr before quenching, and hence SMGs would not be
being observed at a particularly special time.  The latter
interpretation is more consistent with our scenario that SMGs are
not typically undergoing a spectacularly life-changing event.  We note
that there would still be some phenomenon required to quench 
star formation in massive galaxies to produce the red and dead systems seen
today, but perhaps it does not have to be directly associated with the
SMG phase.

\section{Conclusions}\label{sec:summary}

We investigate the physical properties of the most rapidly star-forming
galaxies at $z=2$, constrained to match the observed number density
of sub-millimetre galaxies, in a $\Lambda$CDM hydrodynamic simulation
of galaxy formation.  We examine their stellar masses, star formation
rates, metallicities, environments, clustering, star formation histories,
morphologies, and kinematics.  We find that the simulated high-SFR
galaxies are a good match to the observed SMGs in terms of their stellar
masses, number densities, and clustering, which is a first for a
cosmologically-situated galaxy formation model.

However, there is a significant discrepancy between the predicted and
observed star formation rates:  The simulated SMG's SFRs are lower by a
factor of $\sim 3$ compared to that inferred for SMGs from their observed
far-IR fluxes.  We argue that it is plausible that the SFRs of SMGs have
been overestimated by this factor.  There is growing evidence that such
a modest overestimate would reconcile various observations related to
star formation and stellar mass growth at $z\sim 2$.  Given that high-$z$
galaxies are in many ways systematically different than present-day ones
(e.g. they have much higher star formation rate surface densities),
it is possible that the conversion factors calibrated to local galaxies
may be incorrect at high-$z$.  One way to alter such calibrations would
be to invoke an IMF with more massive stars relative to low-mass ones.
This IMF variation is within currently allowed constraints, and is far
milder than the extremely top-heavy IMF proposed in the \citet{bau05}
semi-analytic models.  However, there are many other possible reasons
why the SFR calibration could be off for SMGs.

We note that the agreement with the observed stellar masses depends with which
observational sample one compares.  Most analyses seem to favour large
stellar masses for the SMGs, typically $>10^{11} M_\odot$, but some obtain
significantly lower stellar masses, even using identical data.
This illustrates the current uncertainties in determining stellar masses
from photometric data, especially in the presence of large amounts of
extinction and possible AGN contamination, in a regime that is poorly
calibrated by local data.  If the lower stellar masses as inferred by e.g.
\citet{hai08} are confirmed, our model would be significantly less
preferred, as it would require an order of magnitude overestimate of the 
observed star formation rates to reconcile our simulations with such data.
Furthermore, it would alleviate one of the difficulties with the major
merger model, namely that large major mergers are too rare, since these
lower-mass systems are more common.  However, it is still true that
the merger model would have to explain the enormous far-IR fluxes,
which seem to require quite massive gas-rich progenitors or else a
dramatic change in the IMF.

If our scenario is correct, the most significant implication would be
that SMGs are typically not large major mergers of gas-rich disks, as
is canonically believed.  Hence, they are in this sense not analogs of
local ULIRGs, in spite of their high bolometric fluxes.  They are instead,
to first order, super-sized versions of ordinary star forming galaxies,
sitting at the high-mass end of the star formation rate--stellar mass
relation.  They are being supplied with gas from minor mergers and smooth
cold accretion at a rate comparable to their SFR.  The SMG phase is,
therefore, not a particularly dramatic event in the life of a massive
galaxy, although their star formation must eventually be quenched by some
mechanism for these systems to become brightest group ellipticals today.

It is worth noting that our simulations do not produce a population
of passively evolving galaxies as observed, likely because we do
not explicitly include any physical process that fully quenches star
formation, such as AGN feedback~\citep[e.g.][]{dim08}.  Passively evolving
galaxies are seen at $z\sim 2$~\citep[e.g.][]{kri08}, and at the massive
end perhaps as many as half the galaxies are passively evolving, or
at least forming stars at a reduced rate for their mass~\citep{pap06}.
If we declared half our massive galaxies to be passively evolving, then to
reproduce the number density of SMGs, we would have to go to a somewhat
lower SFR threshold of $120\; M_\odot/$yr (vs. $180\; M_\odot/$yr).
Hence, to match the observed SFRs we would then require a reduction
by a factor of $\sim 4$ in SFR instead by a factor of $\sim 3$ owing
to some systematic, which is still not implausible.  Of course a full
galaxy formation model should account for passively evolving systems,
but the fact that our current simulations do not is unlikely to affect
our basic conclusions here.

Although SMGs are not particularly distinguished from the more
typical galaxy population other than by their larger stellar masses,
they are still effectively a star formation rate-selected sample.
Therefore, we predict that there will be mild trends to higher SFRs,
lower metallicities, and higher gas fractions than for non-SMGs at
comparable stellar masses.  In other words, the specific star formation
rate of SMGs will be higher than the average at their stellar mass, and
this will be particularly true for lower-mass SMGs owing to the scatter in
the SFR$-M_*$ relation.  Stated in terms of the star formation activity
parameter $\asf\equiv (M_*/{\rm SFR})/(t_H-1\;{\rm Gyr})$, one expects
that SMGs at higher masses will have a higher $\asf$, since they will
be less likely to have a temporarily elevated SFR.  If this trend is
confirmed in future studies (which will require much-improved measures of
SFR and $M_*$) it would provide further support for our overall scenario.

Our model predicts further trends that should become evident with
larger and deeper SMG samples: (i) Lower-$M_*$ SMGs will appear
preferentially more disturbed or merger-like, while high-$M_*$ systems
will appear more quiescent; (ii) Going a factor of a few deeper in
far-IR luminosity should start to suggest a trend in SFR$-M_*$ for SMGs;
(iii) This trend should smoothly join onto the trend of SFR$-M_*$ seen
in optically-selected galaxies; (iv) SMGs should be highly clustered,
and the clustering strength should increase with SMG luminosity;
(v) The occupancy fraction of SMGs within massive halos, as might be
quantified by comparing the clustering and number densities of SMGs,
should be fairly high.  These trends are likely subtle, so testing them
will require improved modeling along with improved data.  We particularly
emphasize the importance of accurate stellar mass and clustering measures
to provide optimal discrimination between models.  Naively, one might
expect that the above trends would not exist or would be weaker in a
major merger-dominated scenario for SMGs, but this must be assessed
within a full hierarchical model.

All the current scenarios for SMGs, i.e. the large major merger scenario,
the modest merger scenario with very top-heavy IMF, and our super-sized
star-forming galaxy scenario, each have their own difficulties.  Our main
point is that the latter scenario is at least as consistent with the
available observations of SMGs as all the others.  Hence, there should
be some doubt as to whether or not SMGs are systems caught in the act
of a major merger.  Even so, major mergers of large galaxies almost
certainly occur, and based upon the simple number density estimates we
presented earlier (\S\ref{sec: intro}), it might be expected that perhaps
one or two in ten SMGs are indeed such systems.  In fact, this may be
the only way to produce the most extremely luminous SMGs, e.g. with
$L_{IR}>10^{13} L_\odot$.  Therefore, we speculate that perhaps SMGs
are a heterogeneous population, and that the very most extreme systems
are indeed large major mergers, while the more common modestly-luminous
SMGs are super-sized star formers.

SMGs are a key phase in the evolution of massive galaxies, possibly
the most active phase of formation for present-day passive spheroids.
Assembling a fully self-consistent cosmological framework for their
formation and evolution will require much effort on both theoretical and
observational fronts.  Fortunately, there will be substantial observational
advances forthcoming in long-wavelength capabilities to trace star formation,
dust continuum emission, and molecular lines of SMGs out to high-$z$.
These will be complemented by better multiwavelength data to constrain
their stellar masses, AGN content, clustering, and metallicities.
Together with rapidly improving models of SMGs within a cosmological
framework, there is growing hope that the nature of these enigmatic
systems and their importance in our overall picture of galaxy formation
will soon be revealed.

 \section*{Acknowledgements}
The simulations used here were run on University of Arizona's SGI
cluster, ice.  We thank Scott Chapman, Kristen Coppin, Reinhard
Genzel, Desika Narayanan, Alex Pope, J.-D. Smith, and Linda Tacconi
for helpful conversations.  We thank Rob Kennicutt and the Institute
of Astronomy for their hospitality during much of the writing of
this paper.  Support for this work was provided by NASA through grant
number HST-AR-10946 from the Space Telescope Science Institute, which is
operated by AURA, Inc. under NASA contract NAS5-26555.  Support for this
work, part of the Spitzer Space Telescope Theoretical Research Program,
was also provided by NASA through a contract issued by the Jet Propulsion
Laboratory, California Institute of Technology under a contract with NASA.
Computing resources were obtained through grant number DMS-0619881 from
the National Science Foundation.

\end{document}